# A new approach for a safe car assistance system

Mejdi Ben Dkhil, Mohamed Neji, Ali Wali, and Adel M. Alimi
REsearch Groups in Intelligent Machines
University of Sfax, National School of Engineers (ENIS)
BP 1173, 3038
Sfax, Tunisia
{mejdi.bendkhil, mohamed.neji, ali.wali, adel.alimi}@ieee.org

*Abstract*— Drowsiness, which is the state when drivers do not have scheduled breaks while traveling long distances, is the main reason behind serious motorway accidents. Accordingly, experts claim that drowsy state is hard to be recognized early enough to prevent serious accidents that may lead even to road deaths. In this work, we propose a new drowsiness state detection system based on physiological signals and eye blinking. An experiment has been directed to justify the utility of the proposed approach. This system uses a smart video camera that takes drivers faces images and supervises the eye blink (open and close); also, it uses the Emotiv EPOC headset to acquire the electroencephalogram (EEG) signals. Eye detection is done by Viola and Jones technique, EEG. Finally, we have chosen the fuzzy logic techniques to classify the EEG signals and eye blinking detection to analyze the results.

*Keywords-Viola and Jones; drowsiness; eye blinking; EEG; fuzzy logic;*

## I. INTRODUCTION

In literature, driving during drowsiness is a cause for the increasing number of road disasters. In this work [1], the authors have shown that driving drowsiness spread from four to six times the danger of making a road accident. Indeed, in agreement with this work [2], driver drowsiness presents about 20% from all accident causes.

The attempt to develop technologies related to the detection or prevention of drowsiness while driving represents a serious challenge in the choose of accident preventing approaches. Due to the chaos that drowsiness may cause on the road, techniques in this field must to be developed in order to contain its drawbacks .Yet, Driver inattention is the result of driving lack of alertness. Accordingly, driver distraction takes place when an object or event takes out a person's concentration out from the driving mission. The difference between driver distraction and driver drowsiness is that the former requires a sudden factor while the last does not require any sudden factor while the last is indicated by an ongoing extraction of concentration from the road and traffic requests. Even though the causes are different, driver drowsiness and driver distraction share the same events which are reduced driving accomplishment, longer response time, and an aggrandized danger of crash commitment.

Lack of sleep, hard work, time of day, and physical tiresome are the four main factors behind drowsiness. Too busy to accomplish their tasks during the day, people do not have time to get enough sleep. As an attempt to stay awake, they may often resort to caffeine or other stimulants. Hence, staying awake for more than two days, the body will not be able take it more which will cause its collapse and negatively affect the person. Scientifically speaking, the human brain is scheduled based on time of day factor. In fact, the brain monitors the body and controls the times when it should be asleep or stay awake. These times are frequently affiliated with sunup and sundown. Thus, extending the time awake negatively affects the body [3]. As far the physical factor is concerned, we may say that there are people who are on medications or have physical ailments which lead to drowsiness state. In addition, being emotionally stressed will direct the driver to get drowsiness speed [4].

So, for this serious risk, we propose to develop a system which controls the driver drowsiness state in a real time, and alerts driver in critical moment when they are been exhausted in the aim of reducing the rate to have an accident.

The rest of the paper is based on three major points as follows. The second section reviews some related works dealing with the eye expressing recognition while the third deals with the experimental results in this work. Finally, the conclusion and some suggests for future intendeds.

## II. RELATED WORKS

In this paper, we are concentrating on recent works to control the drowsiness's state, if car technologies are going to prevent or at least warn of driver drowsiness, what symptoms does the driver give off that can be detected? According to research, there are multiple categories of technologies that can detect driver drowsiness. Those approaches can be divided into three main categories:

- Vehicle based measures: A number of metrics which includes deviations from lane position, movement of the steering wheel, pressure on the acceleration pedal, etc., are always controlled. Yet, any change in these standards refers to the possibility of a significant increase in the driver drowsiness state.

- Behavioral based measures: The behavior of the driver which includes yawning, eye closure, eye blinking, head pose, head movement, etc .can be controlled through a camera and the driver is alarmed if any of these drowsiness symptoms are detected.

- Physiological based measures: The relation between physiological signals ECG (Electrocardiogram) and EOG (Electrooculogram) is identified. Drowsiness can be detected through pulse rate, heart beat and brain information.

In actuality, Drowsiness detection is based on the three parameters mentioned above. In depth study, these measures must be conducted to provide insight on the present systems, issues associated with them and the enhancements that must be implied to make a robust system.

The system consists of two phases: eye detection and EEG analysis.

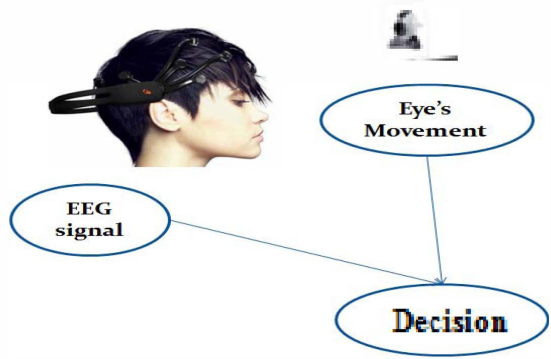

Fig.1. System overview

III. EXPERIMENTAL RESULTS

In this section, we describe different experimental results presents in this work.

We used a video camera and an EEG headset in order to determine the drowsiness state. For acquiring the EEG Data, we have used the Emotiv EPOC headset containing 14 electrodes (AF3, F7, F3, FC5, T7, P7, O1, O2, P8, T8, FC6, F4, F8 and AF4) and 2 references electrodes. These electrodes are placed according to the standard layout of the 10-20 system. This system refers to the spacing of the electrodes varying from 10 or 20% according to the morphology of the individual (Fig2.).

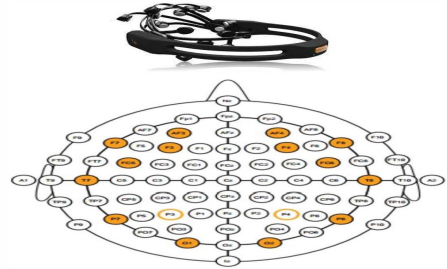

Fig.2. Emotiv EPOC headset

A. *System Flowchart*

The system architecture flowchart is shown in Fig.3.

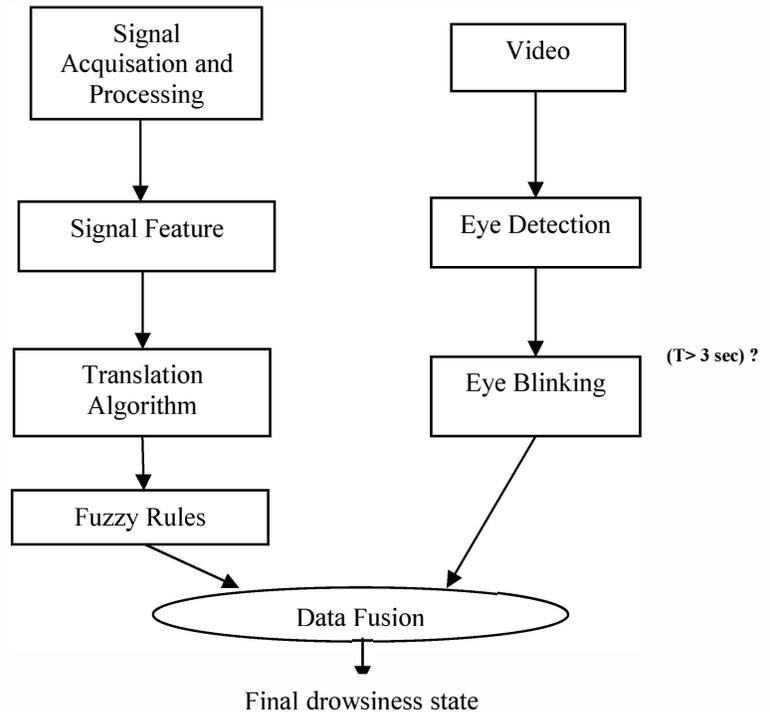

Fig.3. Proposed approach for detecting driver drowsiness

B. *Eye detection and blinking*

Face and eyes are distinguished by the technique for Viola-Jones. This strategy permits the location of items for which learning was performed [5, 6]. It was composed particularly with the end goal of face location, yet might likewise be utilized for different sorts of articles. As an administered learning system, the strategy for Viola-Jones obliges hundreds to a great many samples of the located item to prepare a classifier. The classifier is then utilized as a part of a comprehensive quest of the item for all conceivable positions and sizes the image to be prepared [7].

This system has the playing point of being compelling, fast. The system for Viola-Jones utilizes manufactured representations of pixel values: the pseudo-Haar characteristics. These attributes are controlled by the distinction of wholes of pixels of two or more contiguous rectangular areas (Fig 4.), for all positions in all scales and in a detection window. The number of features may then be high. The best peculiarities are then chosen by a technique for boosting, which gives a "solid" classifier all the more by weighting classifiers "weak".

The Viola-Jones method used by the Adaboost algorithm.

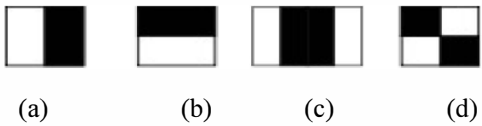

(a)　　　(b)　　　(c)　　　(d)

Fig.4. Examples of neighborhoods used.

The exhaustive search for an item inside an image which can be measured in computing time. Every classifier decides the vicinity or nonappearance of the item in the image. The least difficult and quickest classifiers are put in the first place, which rapidly disposes of numerous negative.

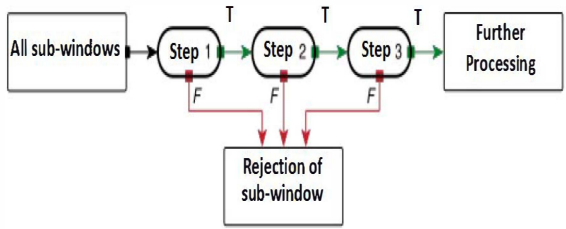

T: True
F: False

Fig.5. Cascade of classifiers [8].

In general, the technique for Viola-Jones gives great results in the Face Detection or different articles, with few false positives for a figuring time low, permitting the operation here progressively [9].

The recognition of eye squinting progressively is essential to gauge driver tiredness state.

In writing, the PERCLOS (PERcentage of eye CLOSure) esteem has been utilized as laziness metric which demonstrates the rate of time in a moment that the eyes are 80% shut [10]. Utilizing these eyes closer and flickering proportion one able to discover laziness of the driver. We move to the accompanying casing until getting closed eyes.

We calculate the duration of eye closure; if it exceeds a preened Time T (for example 3 seconds), we can say that the driver enter in drowsiness, so our system should releases an alert to wake him up. The Theta, Alpha and Beta sub-groups are of enthusiasm for drowsy driving identification [11]. Alpha activity is predominant when an individual is resting or shutting his eyes. Amid the move from alert to rest or sluggish state, sleepiness data can be passed on by an extensive variety of multimodal data signals. In drowsiness location for Human Computer Interaction (HCI), the greater part of the signs can be consolidated into four sorts.

### C. Electroencephalographic signals

As a matter of fact, the physiological signal most commonly used to measure drowsiness is called the Electroencephalogram (EEG). The EEG signal has numerous frequency bands, including the delta band that it is related to sleep activity, the theta band which corresponds to drowsiness, the alpha band representing relaxation and creativity, and the beta band associated to alertness. Drowsiness is indicated by a decrease in the power in the alpha and an increase in the theta frequency [12].

TABLE I. EEG RHYTHMS

| Rhythms | Frequency interval | Location | Reason |
|---|---|---|---|
| Delta | (0-4) Hz | Frontal lobe | Deep sleep |
| Theta | (4-7) Hz | Median, temporal | Drowsiness and meditation |
| Alpha | (8-13) Hz | Frontal, occipital | Relaxation and closed eyes |
| Mu | (8-12) Hz | Central | Controlateral and motor acts |
| Beta | (13-30) Hz | Frontal, central | Concentration and reflection |
| Gamma | (>30) Hz | — | Cognitive functions |

According to Table I, we conclude that The Theta activity is interested for drowsy driving identification and the Alpha activity is predominant when a person is resting or closing his eyes. During the transition from awake to sleep or drowsy state, drowsiness information can be conveyed by a wide-range of multimodal information signals. In drowsiness detection for Human Computer Interaction (HCI), all of the signals can be combined into four types.

1. Feature extraction

- Arousal: It is characterized by a high beta power and coherence in the parietal lobe as well as low an alpha activity. Beta waves are associated with a state of an alert or excited mind, while alpha waves are more dominant in a state of relaxation. Thus the beta/alpha ratio is a reasonable indicator of the excitation state of a person [13].

$$Arousal = [α(AF3 + AF4 + F3 + F4)]/[β(AF3 + AF4 + F3 + F4)] \quad (1)$$

- Valence: The prefrontal lobe (F3 and F4) plays a crucial role in the regulation of drowsiness and conscious experience.

$$Valence = αF4/βF4 - αF3/βF3 \quad (2)$$

- Dominance: It is characterized by an increase in the ratio beta / alpha activity in the frontal lobe and an increase in the beta activity in the parietal lobe.

$$Dominance = (βFC6/αFC6) + (βF8/αF8) + (βP8/αP8) \quad (3)$$

To extract the arousal, valence and dominance drowsiness features, we present the following algorithm [14]:
a) Loading the CSV file.
b) For each data of the signal: (treatment of 20 s of the signal).
- Applying the FFT filter on the signal of electrodes
- Construction of the two pass-band filter Alpha and Beta.
- Applying the pass-band filter on the different electrodes.
- Computing the values of Arousal, valence and dominance ( given in (1), (2) and (3)).

End For.
c) Clustering by Fuzzy Cmeans.

2. Fuzzy logic classification

In this section, we resort to classify the EEG signals. We use the fuzzy logic techniques, among which the Mamdani one is used (Fig.6).

We have three inputs: arousal, valence, dominance and the state of drowsiness is the output variable.

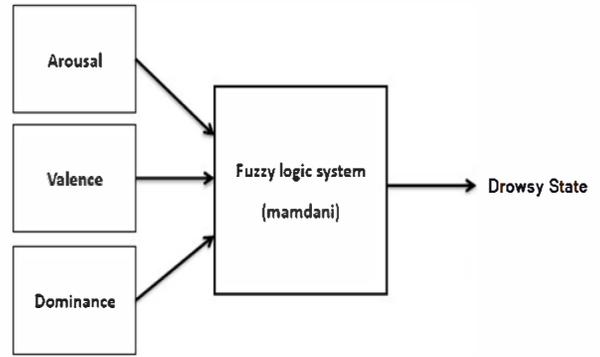

Fig.6. Fuzzy logic system

Each input variable has three membership functions: Small, Medium and Large.

We present in what follows the fuzzy rules:

1) if (arousal is Medium) then (drowsiness is Small)

2) if [(arousal is Small) and (valence is Small) and (dominance is Small)] then (Drowsiness is Small)

3) if [(arousal is Large) and (valence is Large) and (dominance is Large) ] then (Drowsiness is Large)

4) if [(arousal is Large) and (valence is Small) and (dominance is Medium)] then (Drowsiness is Small)

5) if [(arousal is Large) and (valence is Small) and (dominance is Large)] then (drowsiness is Small)

6) if [(arousal is Small) and (valence is Medium) and (dominance is Medium)] then (drowsiness is Small)

7) if [(arousal is Small) and (valence is Large) and (dominance is Medium)] then (drowsiness is Medium)

8) if [(arousal is Small) and (valence is Medium) and (dominance is Large)] then (drowsiness is Small)

9) if [(arousal is Small) and (valence is Large) and (dominance is Large)] then (drowsiness is Medium)

3. Analysis Process

To execute the fuzzy classifier, we use the "Fuzzy logic controller" included in the Simulink library (Fig.7.). In this application, the Simulink bloc of the EEG signals analysis contains a Matlab function (Function1). The role of this function is to retrieve the file containing the EEG signal, perform the FFT on this signal, apply a pass-band filtering to extract the alpha and beta waves of the concerned electrodes, and then extract the features of the signal by calculating the arousal, valence and dominance values [15].

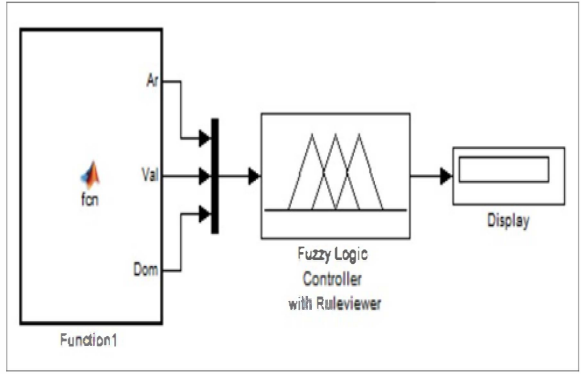

Fig.7. Simulink bloc of EEG signals analysis

The three outputs of this function will be the inputs to the fuzzy logic controller, which will classify the features of the generated signal and generate the drowsiness of the test subject.

*D. Tests and results*

It is very difficult to find databases that merging psychological signals and eye movements. So, in order to validate our proposed system, we have created a set of test samples in our research group.

In first case, we have estimated the drowsiness state by detecting eye blinking (if T>3seconds), this first system has been improved when it was tested with BioIDFaces database [16]. We used the viola and Jones technique to determinate eye detection and blinking.

In second case we developed a set of samples in order to detect drowsiness by analysis EEG signals. We use fuzzy rules defined in section (III.B.2)

Finally, we try to validate our proposed system that controls drowsiness level by analysis both EEG signals and eye blinking. We note:

1.a) if(T=0) then Video=0.

1.b) if ( 0<T<3) then video = 1.

1.c) else video=2.

2.a) if( EEG output = Large) then EEG=2.

2.b) if(EEG output = Medium) then EEG =1.

2.c) else EEG=0.

We define fuzzy Rules:

1) if ((EEG = 2) or (Video =2)) then Drowsiness=2.

2) if (( EEG = 1) and (Video=2)) then Drowsiness=2.

3) if ((EEG=1) and ( Video =1)) then Drowsiness = 1.

4) if ((EEG =0) and (Video =1)) then Drowsiness = 1.

5) if ( (EEG=0) and ( Video = 0)) then Drowsiness =0.

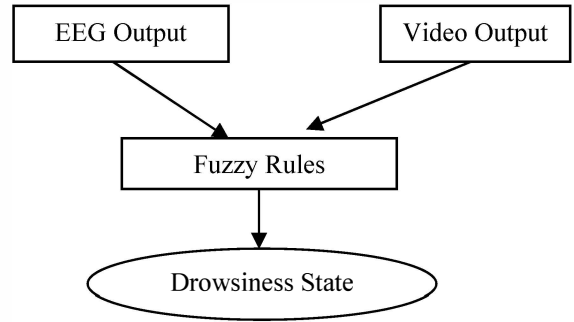

Fig.8. Proposed Fusion System

As a result of the analysis phase, we have obtained a rate of recognition favorable for the drowsiness detection. The rate of recognition is acceptable and susceptible to be improved (Table II).

Table II. DROWSINESS RECOGNITION RATES

| Drowsiness Technique | Recognition rate |
|---|---|
| Eye Blinking | 94.8% |
| EEG signals | 62.14% |
| Eye Blinking and EEG Signals | 70.31% |

IV. CONCLUSION AND FUTURE WORKS

Throughout this work, we have exposed a new system for driver monitoring attention state. As well, this work offers a new system for controlling driver drowsiness state by analyzing of the EEG signals (moral state) and eye blinking (physical state). This developed system is based on computer vision techniques.

As perspectives, we are seeking to propose a safety car assistance system that controls both: the inside car risks (the driver vigilant state) and the outside car risks (pedestrian, moving object, road lanes, and panel roads).

V. ACKNOWLEDGEMENT

The authors would like to acknowledge the financial support of this work by grants from the General Direction of Scientific Research (DGRST), Tunisia, under the ARUB program.